\documentclass[sigconf]{acmart}

\usepackage{svg}
\usepackage{graphicx}
\usepackage{subcaption}
\usepackage{url}
\usepackage{svg}
\copyrightyear{2026}
\acmYear{2026}
\setcopyright{none}
\renewcommand\footnotetextcopyrightpermission[1]{}
\acmDOI{10.1145/3786579.3804952}
\acmISBN{979-8-4007-2422-0/2026/07}
\begin{document}

\title{AktivTalk: Digitizing the Talk Test for Voice-Based Exercise Intensity Self-Assessment and Exploring Automated Classification from Speech}

\author{Rania Islambouli}
\orcid{0000-0003-1689-8938}
\affiliation{%
  \institution{Ludwig Boltzmann Institute for Digital Health and Prevention}
  \country{Austria}
  }
\email{rania.islambouli@lbg.ac.at}

\author{Laura Geiger}
\orcid{0009-0004-7515-7325}
\affiliation{%
  \institution{University of Innsbruck}
  \country{Austria}
  }
\email{lauraxgeiger@icloud.com}

\author{Daniela Wurhofer}
\orcid{0000-0001-8476-6051}
\affiliation{%
  \institution{Ludwig Boltzmann Institute for Digital Health and Prevention}
  \country{Austria}
  }
\email{daniela.wurhofer@lbg.ac.at}

\author{Devender Kumar}
\orcid{0000-0002-6971-2829}
\affiliation{%
  \institution{ University of Southern Denmark}
  \country{Denmark}
  }
\email{deku@mmmi.sdu.dk}

\author{Clemens Sauerwein}
\orcid{0009-0009-9464-5080}
\affiliation{%
  \institution{University of Innsbruck}
  \country{Austria}
  }
\email{Clemens.Sauerwein@uibk.ac.at}

\author{Jan David Smeddinck}
\orcid{0000-0003-0562-8473}
\affiliation{%
  \institution{Ludwig Boltzmann Institute for Digital Health and Prevention}
  \country{Austria}
  }
\email{jan.smeddinck@lbg.ac.at}

\begin{abstract}
Monitoring exercise intensity is critical for safe and effective physical activity, particularly for individuals with cardiovascular disease, where overexertion can pose serious risks. Although physiological measures such as heart rate are widely used for avoiding overexertion, they can be unreliable in certain cases, such as when affected by medication or when wearables are worn too loosely. We introduce \textit{AktivTalk}, a mobile prototype that digitizes the clinically validated Talk Test to support voice-based, in-the-moment self-assessment of exertion. In a within-subject study with 20 participants, we collected exertion-labeled voice samples and found that \textit{AktivTalk} was rated as highly usable and preferred over conductor-guided assessment. We further explored automated exertion classification from Talk Test speech. Using MFCC-based features with class balancing and cross-validation, a lightweight neural classifier achieved up to 90\% accuracy for detecting \textit{high} vs.\textit{non-high} exertion from Talk Test recordings. This work-in-progress highlights the potential of structured voice interactions for accessible exertion assessment and motivates future passive exertion monitoring from speech.
\end{abstract}



\begin{teaserfigure}
\centering
  \includegraphics[width=0.9\textwidth]{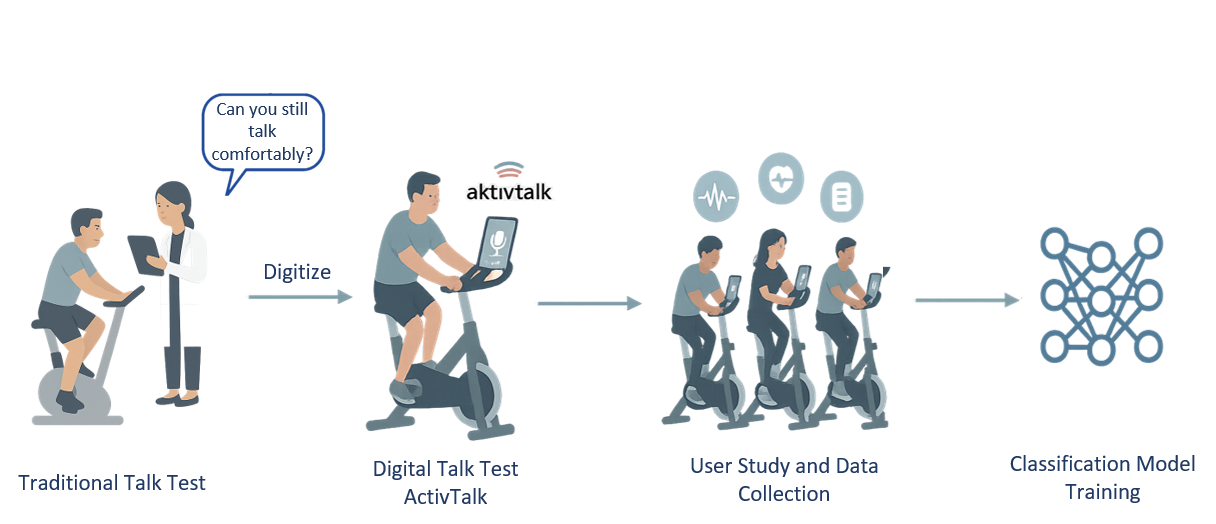}
  \caption{From the traditional Talk Test (left), \textit{AktivTalk} enables digital, voice-based exercise intensity self-assessment. Data collected in a user study (middle) is used to train a machine learning model (right) for automatic classification of exercise intensity.}
  \label{fig:teaser}
\end{teaserfigure}

\maketitle
\section{Motivation and Related Work}
\label{sec:motivation}

Monitoring exercise intensity supports safety and effectiveness in physical activity, particularly in rehabilitation contexts where overexertion can be risky~\cite{hansen2022exercise}. Heart-rate based monitoring is widely used, but can be unreliable when the chronotropic response is attenuated (e.g., beta-blockers) or when wearables are unavailable or noisy due to fit and motion artifacts~\cite{achten2003heart,tesch1985exercise,stahl2016accurate,gillinov2017variable,islambouli2025towards}. These constraints motivate low-barrier alternatives that remain usable during activity. The clinically validated \textit{Talk Test} provides an equipment-free approach grounded in the ability to speak during exertion: users read a short standardized passage and report whether they can still speak comfortably, yielding light/moderate/high intensity zones that align with ventilatory thresholds~\cite{reed2014talk,persinger2004consistency,quinn2011talk}. Despite its simplicity, the Talk Test is rarely operationalized in consumer-facing tools for unsupervised use, limiting structured guidance, logging, and feedback outside supervised settings~\cite{reed2014talk,vieira2022application}. Separately, microphones in phones and wearables have enabled audio-based health sensing, including speech markers for neurological and respiratory conditions~\cite{ijaz2022towards,donaghy2024review,lentz2025voice}. In the exercise domain, prior work has estimated exertion from breathing audio or speech, typically using physiological ground truth (e.g., heart rate) to define intensity zones~\cite{ren2021breathing,zhou2025voice}. However, there is limited work on modeling \textit{subjective} exertion from speech elicited through a clinically meaningful, structured protocol such as the Talk Test.

We present \textit{AktivTalk}\footnote{Source code: \url{https://github.com/LBI-DHP/ActivTalk}}, a mobile prototype that digitizes the Talk Test as a guided, voice-based self-assessment interaction and logs exertion-labeled speech during exercise. Figure~\ref{fig:teaser} summarizes our approach: digitizing the Talk Test into a mobile app, evaluating usability in a user study, collecting exertion-labeled speech, and exploring voice-based intensity classification. We evaluate usability and preference in a within-subject user study (N=20) and explore feasibility of automatic classification from speech, focusing on detecting transitions into \textit{high} exertion due to its safety relevance in rehabilitation~\cite{milani2024exercise}. Our contributions are: (1) a self-guided Talk Test mobile prototype; (2) comparative study findings on usability, trust, and preference across administration modes; and (3) initial evidence that speech features can distinguish high vs.\ non-high exertion.

\section{AktivTalk Prototype}
\label{sect:design}

\begin{figure*}[t]
     \centering
  \hspace*{-3em}
\includegraphics[width=0.85\textwidth]{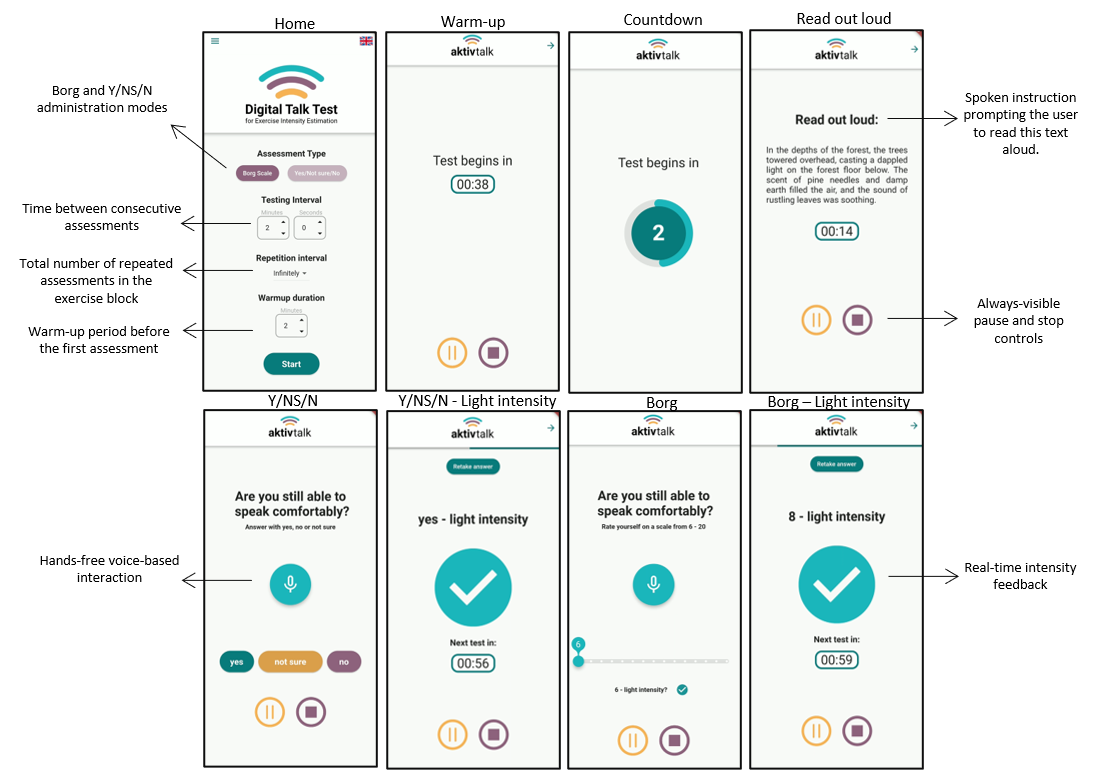}
    \caption{\textit{AktivTalk} UI workflow, showing user flow from configuration through repeated assessment cycles with audio capture, self-rating, and feedback.}
    \label{fig:ui_workflow}
\end{figure*}

\textit{AktivTalk} digitizes the clinically validated Talk Test as a self-guided, voice-based interaction that can be performed during exercise without supervision or external equipment. The design preserves the core protocol of the Talk Test (reading a standardized passage followed by self-assessment) while remaining usable under physical exertion, where attention and motor precision are constrained. Design requirements were informed by interviews with two HCI researchers and two sports scientists.  These expert inputs helped us identify three core functions: the app (1) must preserve clinical validity and enable digital capture; (2) must support reliable interaction during exertion with minimal motor demand and error recovery; and (3) must support flexible assessment by offering both categorical Yes/Not sure/No (YNNS) and BORG RPE (6-20)~\cite{bok2022examination} (for a more fine-grained scale for self-assessment ) input formats. Figure~\ref{fig:ui_workflow}  illustrates the AktivTalk workflow. Users configure an exercise session (method, interval, repetitions, warm-up) and complete repeated assessment cycles consisting of reading a standardized 30-word paragraph while audio is recorded, providing an exertion rating, and receiving immediate feedback. YNNS responses map directly to intensity zones (Yes=light, Not sure=moderate, No=high). BORG ratings are mapped to the same three zones using Talk Test staging: 6-10 (light), 11-13 (moderate), and $\geq$14 (high)~\cite{bok2022examination}. To support use during movement, AktivTalk uses large, high-contrast controls, minimal navigation depth, and a persistent \texttt{Pause} action. Input is multimodal to also enable hands-free interaction: voice input is the default, with automatic fallback to touch controls when no confident speech is detected within a short timeout. A brief retake window supports recovery from misrecognition or accidental taps. AktivTalk was implemented in Flutter/Dart to support cross-platform deployment. For the study, recordings were stored locally as 15-second WAV files with timestamps and session metadata; each assessment logged the selected rating, method, and associated audio file. The app operates offline and supports later export.

\section{User Study}
\subsection{Study Design}
\begin{figure*}[t]
    \centering
    \includegraphics[scale=0.41]{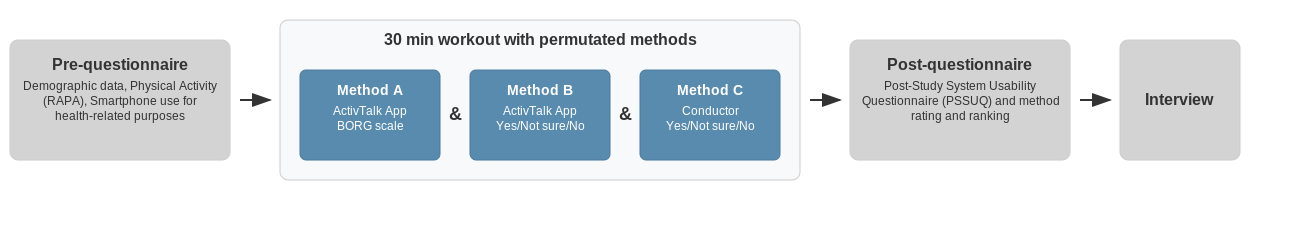}
    \caption{Overview of the user study protocol showing the sequential phases: pre-study questionnaire, 30-minute workout with permuted methods, post-study questionnaire, and semi-structured interviews.}
    \label{fig:study_pt}
\end{figure*}

We conducted a within-subject study to evaluate the usability, trust, and user experience of \textit{AktivTalk}, compare Talk Test administration methods, and collect exertion-labeled speech for exploratory modeling. We recruited 20 participants (10 female, 10 male; age 22-58, M=34.1, SD=8.3) via snowball sampling. All were experienced smartphone users; individuals with known cardiovascular or respiratory conditions were excluded. Physical activity levels were assessed using RAPA~\cite{topolski2006rapid}. 11 participants were classified as active, 3 as under-active regular, 4 as
under-active regular with light activities, and 2 as under-active. The study was approved by the organizational ethics committee (details anonymized). Sessions were conducted indoors on a stationary spin bike (Figure~\ref{fig:study_pt}). Each participant completed three 10-minute cycling blocks in permuted order: (A) \textit{Digital-BORG} using AktivTalk with BORG RPE input; (B) \textit{Digital-YNNS} using AktivTalk with YNNS input; and (C) \textit{Conductor-YNNS}, where a conductor administered the Talk Test without the app. A Talk Test occurred every minute. Method order was counterbalanced via a Latin square. Audio was recorded via the app, and heart rate was recorded continuously via a Polar OH1+ sensor. After the session, participants completed the PSSUQ~\cite{lewis2002psychometric}, rated ease of use for each method (1=very easy, 5=very hard), ranked methods by preference, and participated in a brief semi-structured interview about usability, trust, and comfort.

\begin{figure}[t]
    \centering
    \includegraphics[width=0.5\textwidth]{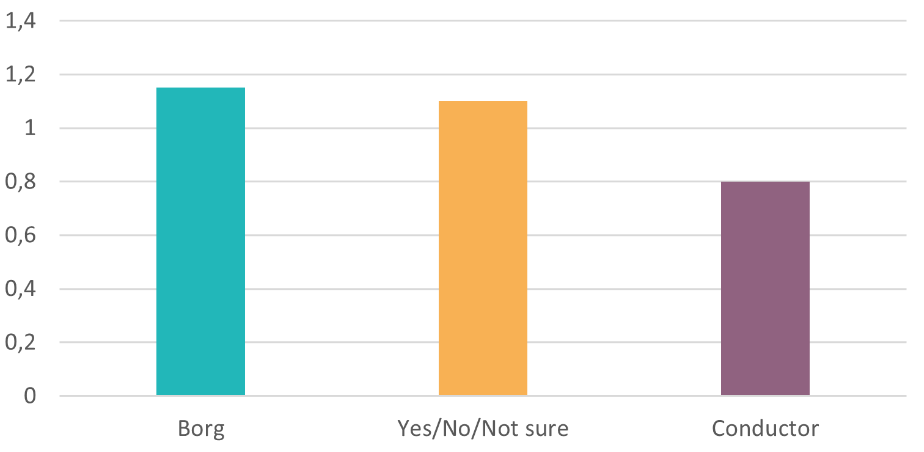}
    \caption{Average participant rankings of the three assessment methods. Higher scores indicate greater preference (2 = most preferred, 0 = least).}
    \label{fig:ranking}
\end{figure} 
\subsection{Results}
\begin{figure*}[t]
    \centering
    \begin{subfigure}[b]{0.29\textwidth}
        \includegraphics[width=\textwidth]{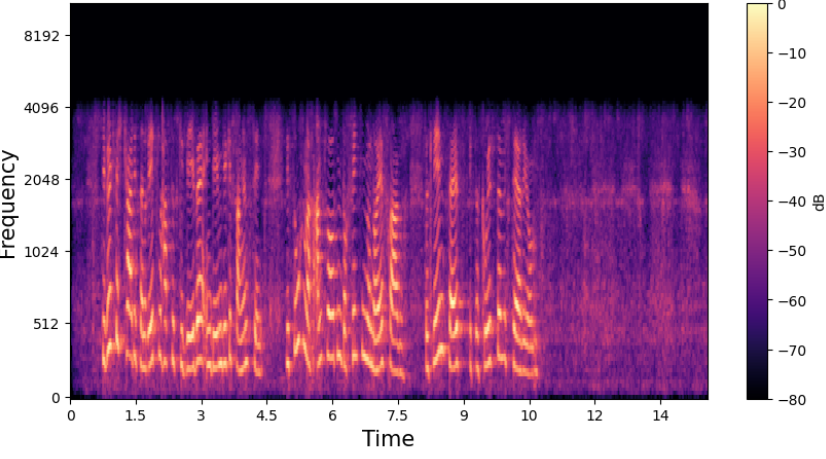}
        \caption{Light}
        \label{fig:spectrogram_light}
    \end{subfigure}
    \hfill
    \begin{subfigure}[b]{0.29\textwidth}
        \includegraphics[width=\textwidth]{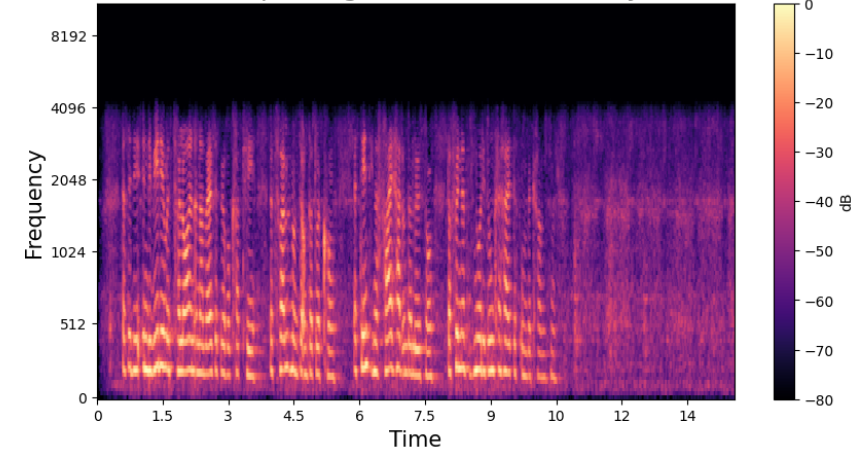}
        \caption{Moderate}
        \label{fig:spectrogram_moderate}
    \end{subfigure}
    \hfill
    \begin{subfigure}[b]{0.29\textwidth}
        \includegraphics[width=\textwidth]{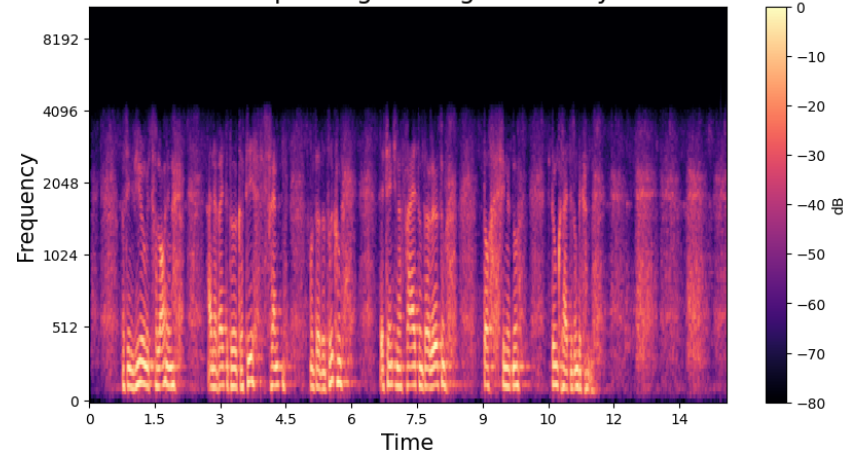}
        \caption{High}
        \label{fig:spectrogram_high}
    \end{subfigure}
    \caption{Mel-spectrograms of speech samples at different exertion levels. Brighter areas indicate higher energy; breathing pauses and irregularities become more pronounced with increasing exertion.}
    \label{fig:spectrograms}
\end{figure*}

\textbf{Usability and ease of use:} AktivTalk was rated highly usable on the PSSUQ (1=best): \textit{System Usefulness} (M=1.28, SD=1.29), \textit{Information Quality} (M=1.58, SD=1.22), and \textit{Interface Quality }(M=1.23, SD=0.64). All methods were rated easy to use (App-YNNS: M=1.55, SD=0.74; Conductor-YNNS: M=1.60, SD=0.86; App-BORG: M=1.65, SD=0.85), with slightly higher interaction effort reported for the BORG slider during high exertion. In interviews, participants described the app as intuitive and glancable during exercise, and suggested larger touch targets and variation in reading text. \textbf{Preference:} Participants preferred app-based Talk Test variants over conductor-guided administration (Figure~\ref{fig:ranking}), commonly citing autonomy and reduced social pressure. \textbf{Speech samples and heart rate: } Across conditions, 559 speech samples were collected and labeled using self-reported exertion. BORG ratings were grouped into light (6-10), moderate (11-13), and high ($\geq$14). YNNS responses mapped to light/moderate/high as Yes/Not sure/No. To relate subjective ratings to physiological effort, samples were synchronized with heart rate data using system timestamps; mean heart rate increased across self-reported intensity levels for all methods (Table~\ref{tab:mean_pulse}). This confirms that users’ self-perceived exertion tracks actual physiological load, a form of sequential validity that supports the core assumptions of the Talk Test~\cite{reed2014talk}. \textbf{Acoustic patterns: } Representative mel-spectrograms (Figure~\ref{fig:spectrograms}) illustrate that high-intensity speech more often exhibits irregular pacing and frequent breathing interruptions, while light and moderate speech show more overlapping patterns.

\begin{table}[t]
\centering
\setlength{\tabcolsep}{2pt} 
\caption{Mean pulse rate (bpm) by assessment method and self-reported intensity}
\label{tab:mean_pulse}
\begin{tabular}{lccc}
\toprule
Assessment Method & Light (bpm) & Moderate (bpm) & High (bpm) \\
\midrule
App-YNNS & 130.21 $\pm$ 20.39 & 151.71 $\pm$ 15.66 & 165.05 $\pm$ 13.98 \\
App-BORG & 126.76 $\pm$ 18.19 & 147.15 $\pm$ 15.79 & 162.51 $\pm$ 12.16 \\
Conductor-YNNS & 131.94 $\pm$ 15.90 & 149.83 $\pm$ 11.17 & 162.14 $\pm$ 12.44 \\
\bottomrule
\end{tabular}
\end{table}

\section{Automatic Intensity Classification (Feasibility)}

Beyond manual self-assessment, we explored whether speech recorded during the Talk Test contains sufficient signal to \textit{automatically} detect exercise intensity. Our goal was not to optimize a production-ready classifier, but to assess feasibility and identify which intensity distinctions are realistically learnable from speech in this context. Given the safety relevance of overexertion in rehabilitation, we focused particularly on detecting transitions into \textit{high} intensity.

\textbf{Dataset and labels.} The dataset comprised 559 Talk Test speech recordings collected during cycling and labeled using participant self-assessments. YNNS responses were mapped to light/moderate/high as Yes/Not sure/No, while BORG ratings were grouped into light (6-10), moderate (11-13), and high ($\geq$14). For comparison, we also derived pulse-based labels from Polar OH1+ recordings using age-predicted maximum heart rate (220$-$age) and standard thresholds~\cite{fox1968physical,seo2021optimal}. This dual labeling allowed us to contrast \textit{subjective} versus \textit{physiological} ground truth. \textbf{Feature extraction and modeling.} From each recording, we extracted Mel-Frequency Cepstral Coefficients (MFCCs), capturing spectral characteristics of speech that are sensitive to vocal effort and breath control. MFCCs were mean-pooled over time to obtain a fixed-length feature vector per sample, emphasizing global vocal characteristics rather than fine-grained phonetic content. We trained a lightweight feedforward neural classifier with class balancing to account for the underrepresentation of high-intensity samples. Model evaluation used a participant-independent stratified 5-fold cross-validation to ensure robustness across participants and intensity distributions, and that all samples from a given participant appeared in only one fold. \textbf{Binary vs.\ multi-class framing.} Visual inspection of mel-spectrograms (Figure~\ref{fig:spectrograms}) revealed substantial acoustic overlap between light and moderate intensity speech, while high-intensity samples consistently showed irregular pacing, increased breathing interruptions, and higher overall energy. Based on this observation and the clinical relevance of detecting unsafe exertion, we evaluated both a three-class (light/moderate/high) and a binary classification task, merging light and moderate into a single \textit{non-high} class.
\begin{figure}[th]
    \centering
    \begin{subfigure}[b]{0.36\textwidth}
        \includegraphics[width=\textwidth]{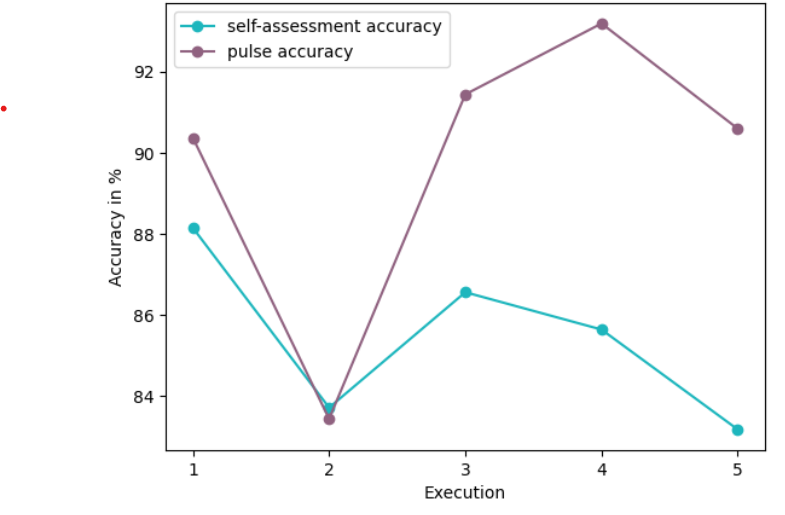}
        \caption{Categorical Classification}
        \label{fig:catg}
    \end{subfigure}
    \begin{subfigure}[b]{0.316\textwidth}
        \includegraphics[width=\textwidth]{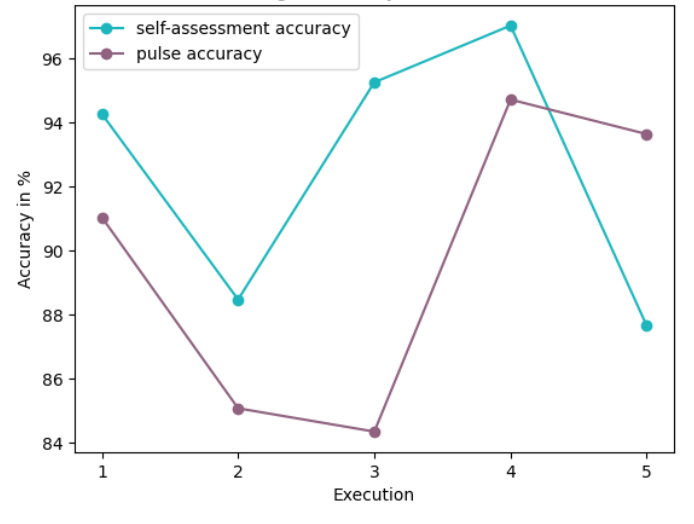}
        \caption{Binary Classification}
        \label{fig:binary}
    \end{subfigure}
    \caption{Model accuracy across 5-fold cross-validation for both classification tasks. (a) Multi-class (categorical) classification: light, moderate, high intensity. (b) Binary classification: non-high vs. high intensity.}
    \label{fig:results_ml}
\end{figure}
\textbf{Results.} In the three-class setting, accuracy averaged $\mathbf{85.44\% \pm 2.09\%}$ for self-assessed labels and  $\mathbf{89.78\% \pm 3.75\%}$ for pulse-based labels over 5-fold 
cross-validation (Figure~\ref{fig:catg}). Performance improved further in the binary classification task. Distinguishing high from non-high exertion achieved an accuracy of 
$\mathbf{92.50\% \pm 4.20\%}$ with self-assessed labels and 
$\mathbf{89.72\% \pm 4.83\%}$ with pulse-based labels over 5-fold cross-validation (Figure~\ref{fig:binary}).

\section{Discussion and Limitations}

AktivTalk demonstrates the feasibility of translating the clinically established Talk Test into a self-guided, voice-first mobile interaction that participants found usable during exertion and preferred over conductor-guided administration. Beyond digitization, our findings show how structured voice interaction can support in-the-moment exertion self-assessment while preserving autonomy. Participants preferred the app-based variants, citing reduced social pressure and greater control compared to conductor-guided assessment. This suggests that digitizing clinician-led protocols may not only increase accessibility but also enhance perceived agency in self-managed exercise contexts. The comparison between YNNS and BORG reveals a granularity–effort tradeoff. While both were rated usable, participants reported slightly greater effort with the BORG slider during high exertion. This suggests that under physiological strain, interaction bandwidth narrows: simple categorical input (YNNS) may better match reduced cognitive and motor capacity, whereas BORG’s finer granularity may be more suitable for reflection or training. Future systems could adapt reporting formats to exertion level, using simple input at peak intensity and finer scales during lower-intensity phases. The automatic intensity classification serves as design evidence that the structured Talk Test interaction produces stable vocal signals for safety-oriented assistance. The clearer separability of high versus non-high exertion suggests that the standardized reading task amplifies exertion-related changes in pacing and breath control, effectively making the interaction itself a sensing instrument. From an HCI perspective, this underscores the value of structured elicitation over passive audio sensing and supports a safety-oriented framing focused on detecting potentially unsafe intensity rather than precise quantification. Beyond feasibility testing, AktivTalk can also function as a data collection instrument, enabling structured, exertion-labeled speech to be gathered in ecologically valid settings and supporting future longitudinal and multimodal modeling work. Limitations should be considered. Data were collected in a controlled indoor cycling setting with healthy adults, limiting generalizability to other activities or clinical populations. The dataset is modest in size, and the model relies on simple aggregated acoustic features; results should therefore be interpreted as feasibility evidence. Pulse-based intensity labels were derived from age-predicted heart rate zoning, which may introduce imprecision. By combining interaction design, user evaluation, and speech-based modeling, AktivTalk illustrates how voice can function both as an
accessible interface and an in the-moment exertion self-assessment during physical activity.

\bibliographystyle{ACM-Reference-Format}
\bibliography{references}

\end{document}